\documentclass[a4paper,12pt]{article}
\pdfoutput=1
\usepackage{mathtext}
\usepackage{misccorr}
\usepackage{indentfirst}
\usepackage{graphicx}
\usepackage{amsmath}
\usepackage{amsbsy}
\usepackage{textcomp}
\usepackage{amssymb}
\usepackage{upgreek}
\usepackage{rotating}
\usepackage{cite}
\usepackage{lineno}
\newcommand {\wrt}{w.r.t. }
\newcommand {\con}{configuration}
\newcommand {\cons} {configurations}
\newcommand {\inz}{inactive zone}
\newcommand {\inzs}{inactive zones}
\newcommand {\COl}{COMET layer}
\newcommand {\ins}{inter-strip}
\newcommand {\inm} {inter-module}
\newcommand {\ine} {inefficiency}
\newcommand {\mst} {muon-stopping target}
\begin{document}
\title{Simulations of the COMET veto counter}
{\sf
\author{Oleg Markin\footnote{e-mail: markin@itep.ru}~ and Evgueny Tarkovsky\\
{\small\emph{Institute for Theoretical and Experimental Physics, 117218 Moscow }}}
\date{\vspace{-5ex}}
\maketitle
\begin{abstract}
A computer model of a scintillator strip veto counter was built in order to verify the efficiency of the cosmic muon veto for the COMET experiment. To tune the model, experimentally measured data were utilized. Three different geometrical configuration of the counter were considered. For one of the configurations the simulation gave the inefficiency of the cosmic muon registration being below 0.0001, which meets requirements  of the experiment.     
\end{abstract}
}
\section{Introduction}
In the COMET experiment the cosmic background should be carefully eliminated  because of tiny number of signal events expected. Veto at a cosmic muon propagation can be effectively managed with scintillator strip plates surrounding the COMET detector \cite{CDR}. Similar strip plates are used in the Belle~II detector where the scintillator light is collected through  Kuraray WLS fibers to Hamamatsu  silicon photodetectors MPPC. Properties of such a detecting system as well as the radiation hardness of MPPC are studied well enough to date \cite{Tar_Vienna,Tar_JNR}. The aim of this study is a computer simulation of inefficiency of the veto counter designed for the COMET. The desirable value of the inefficiency is as small as 0.0001 \cite{CDR}, which can be achieved using a coincident signal of any two from four strip plates in a stack \cite{Tar_JNR,Drutsk}. 

\section{The model}
\label{model}
In this study we only model  a horizontal slab of the COMET veto counter whereas there will be four side slabs also \cite{Drutsk}. Modeling of those vertical slabs is straightforward and probably gives lower inefficiency due to longer average muon range.  The ionization  produced by muon in a scintillator strip is proportional to its range in the strip, hence the signal of muon depends on its entrance position and direction as well as on the size of inactive zones of detector. A layer of the COMET veto counter consists of scintillator strips of 40 mm width and 7 mm thickness. Surfaces of strips are chemically processed to provide a diffuse reflection of the scintillator light. Processing adds 0.2~mm to the strip width on average. Besides, we suppose additional 0.1 mm separation of strips because of imperfect geometry, so two neighboring strips are separated by a 0.3 mm inactive zone.

\begin{figure}
\centering
\includegraphics[width=\textwidth]{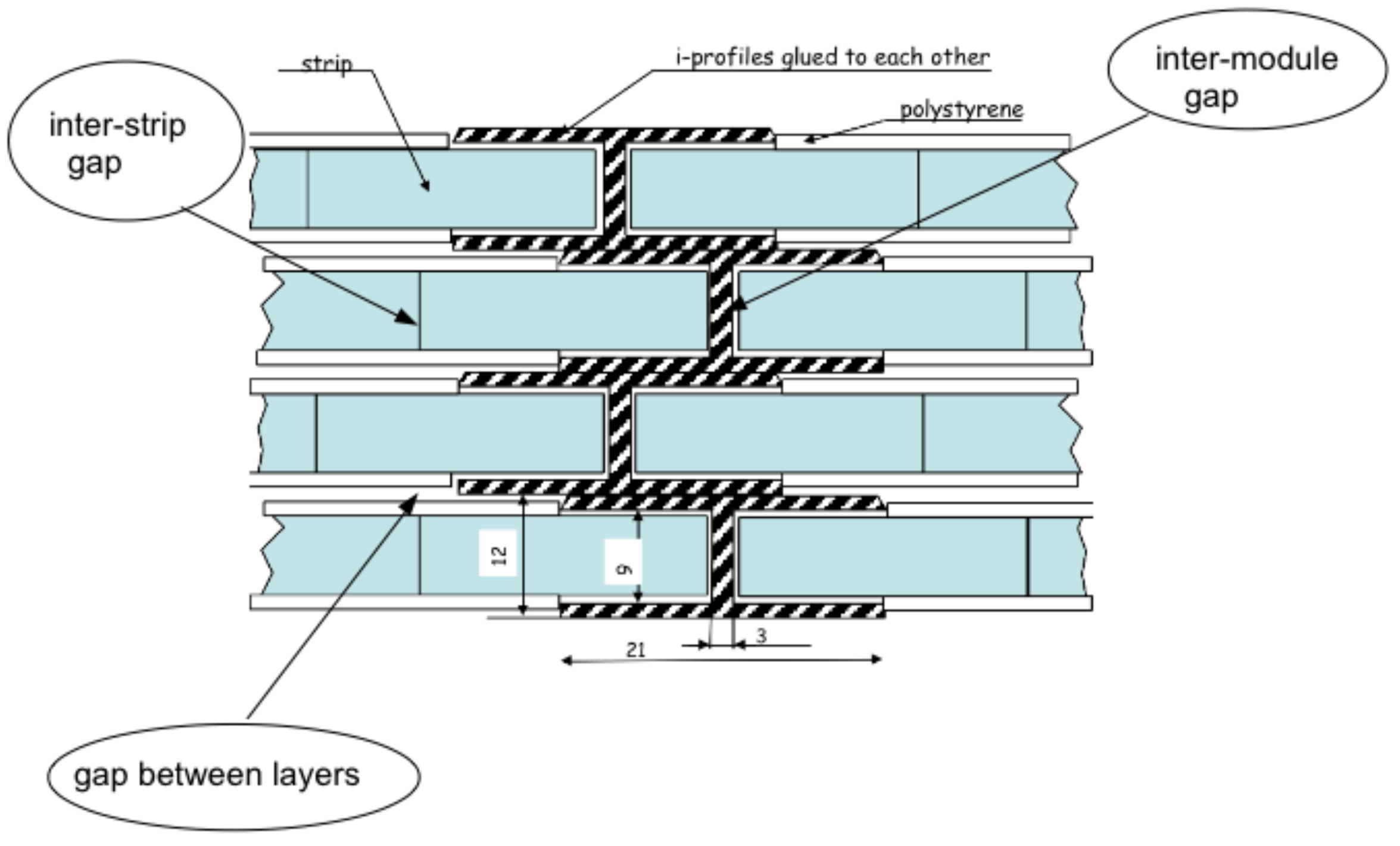}
\caption{ \textsf {One of possible layouts of strips.}}
\label{zones}
\end{figure}

Similar to the Belle~II strip modules, two neighboring modules are separated by an aluminum holder of 5 mm  width, which also is an inactive zone. This wide zone is particularly  harmful for  muon registration because its width is comparable with an average cosmic muon range in  7 mm thick strips, so the muon signal can vanish. To increase the range/signal, both inter-strip and inter-module inactive zones of different layers should be shifted w.r.t. each other, cf. fig.~\ref{zones}.

\begin{figure}
\centering
\includegraphics[width=.8\textwidth]{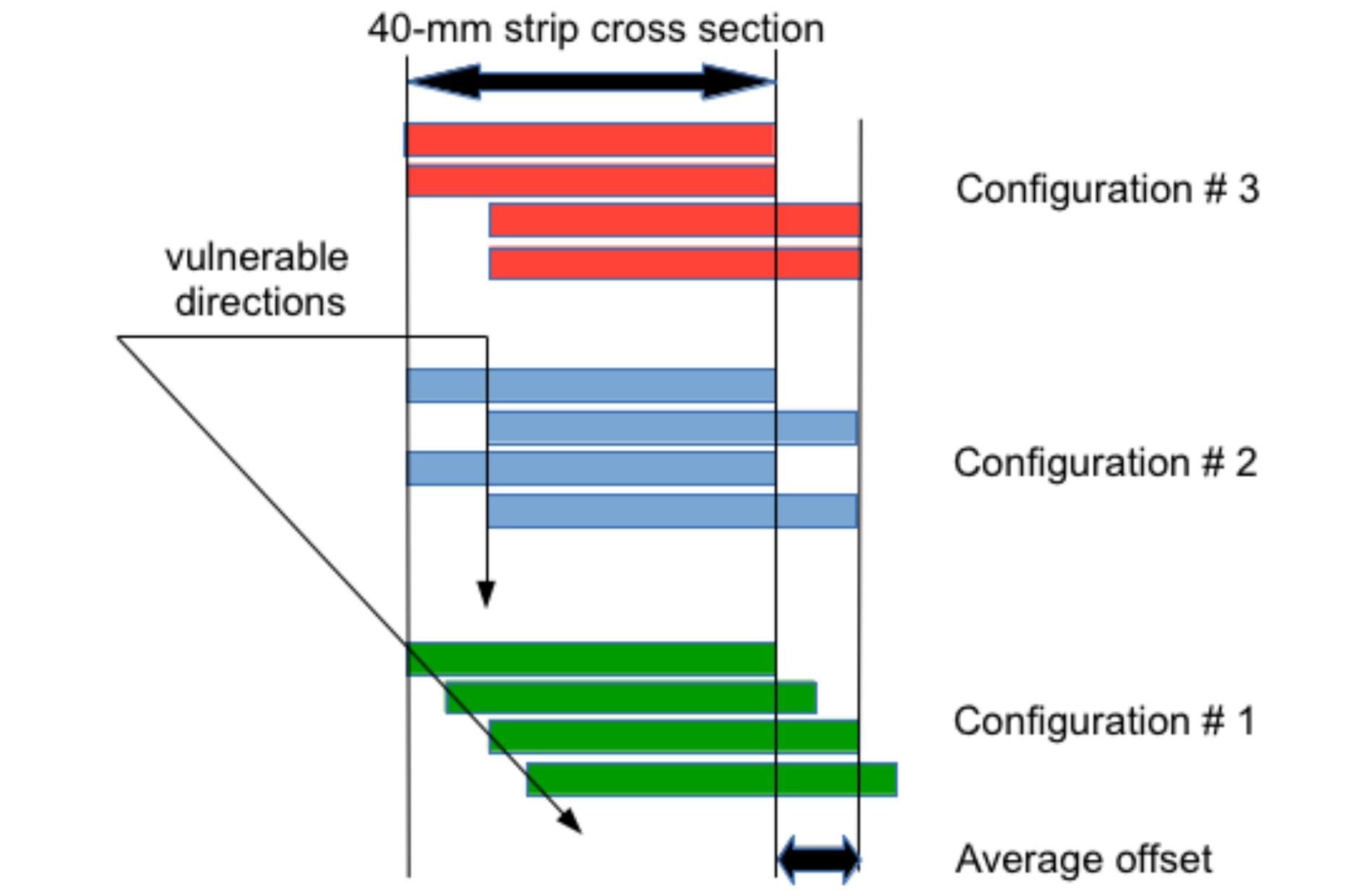}
\caption{ \textsf {Strip layouts implemented in the simulation.}}
\label{conf}
\end{figure}

Our simulations were done for a single COMET layer  as well as for a complete stack of four layers for three configurations sketched in fig.~\ref{conf}. In the first one each subsequent layer is shifted   \wrt  the previous one at the chosen distance in the same transverse direction.  In the second configuration the second and the fourth layers are shifted \wrt  others at the same distance doubled. Finally in the third configuration  the third and the fourth layers are shifted \wrt  others at the same distance doubled. Thus in the latter two configurations the offset of layers coincides with an \emph{average offset} in the the first \con.

\begin{figure}
\centering
\includegraphics[width=.8\textwidth]{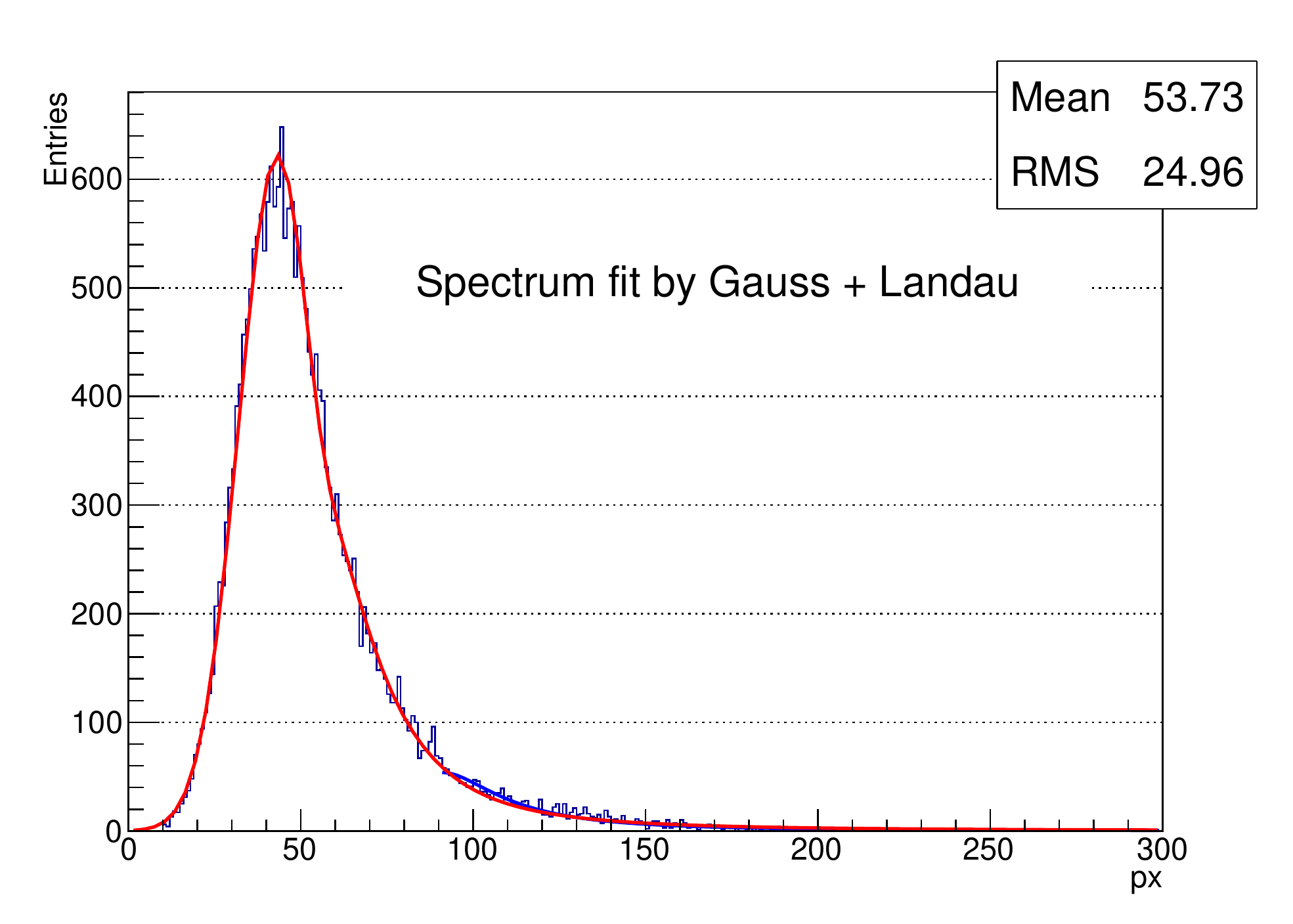}
\caption{ \textsf {The signal spectrum in the middle of a 2218 mm strip.}}
\label{spec}
\end{figure}

On its way from a muon entrance point to the MPPC the collected scintillator light  is significantly attenuated. To take into account the attenuation, experimentally measured spectra were used. The spectra were available for three different locations on 2218 mm strips: in the close part, where a MPPC is set, in the middle and in the rear part of strip, see an example in fig.~\ref{spec}. 

\subsection{The algorithm}
The inefficiency of the four-layer stack was calculated as a fraction  of muons whose signal was below a threshold at least in three layers.\footnote{Alternatively, one can merely count events where the signal in two layers was above the threshold. We have chosen  the former approach to write down the information about muons unregistered in the three layers for a visual verification.}  For this study the threshold value was chosen 11 photo-detector's pixels, which is a possible level of noise signals after a long exposition of MPPC to a neutron background of the experiment, cf. section \ref{noise}. The signal of simulated muon in a strip  was obtained ascribing to each particle  an  experimentally known signal  attenuated and then corrected according to the muon range for given entrance position and direction. To compute the signal, a random number according to a spectrum was taken after the spectrum was interpolated to the entrance position of muon and then scaled  proportionally to the range of given muon in the strip.

Thus the simulation of each event consisted of the following steps:
\begin{itemize}
 \item generating random entrance point of muon within a plate of modules;
 \item generating random polar and random azimuth  angles $\theta$ and $\phi$ of muon trajectory   according to  $\cos^2(\theta)d\cos(\theta)d\phi$  distribution of cosmic muons;
\end {itemize}

\begin{figure}
\centering
\includegraphics[width=.8\textwidth]{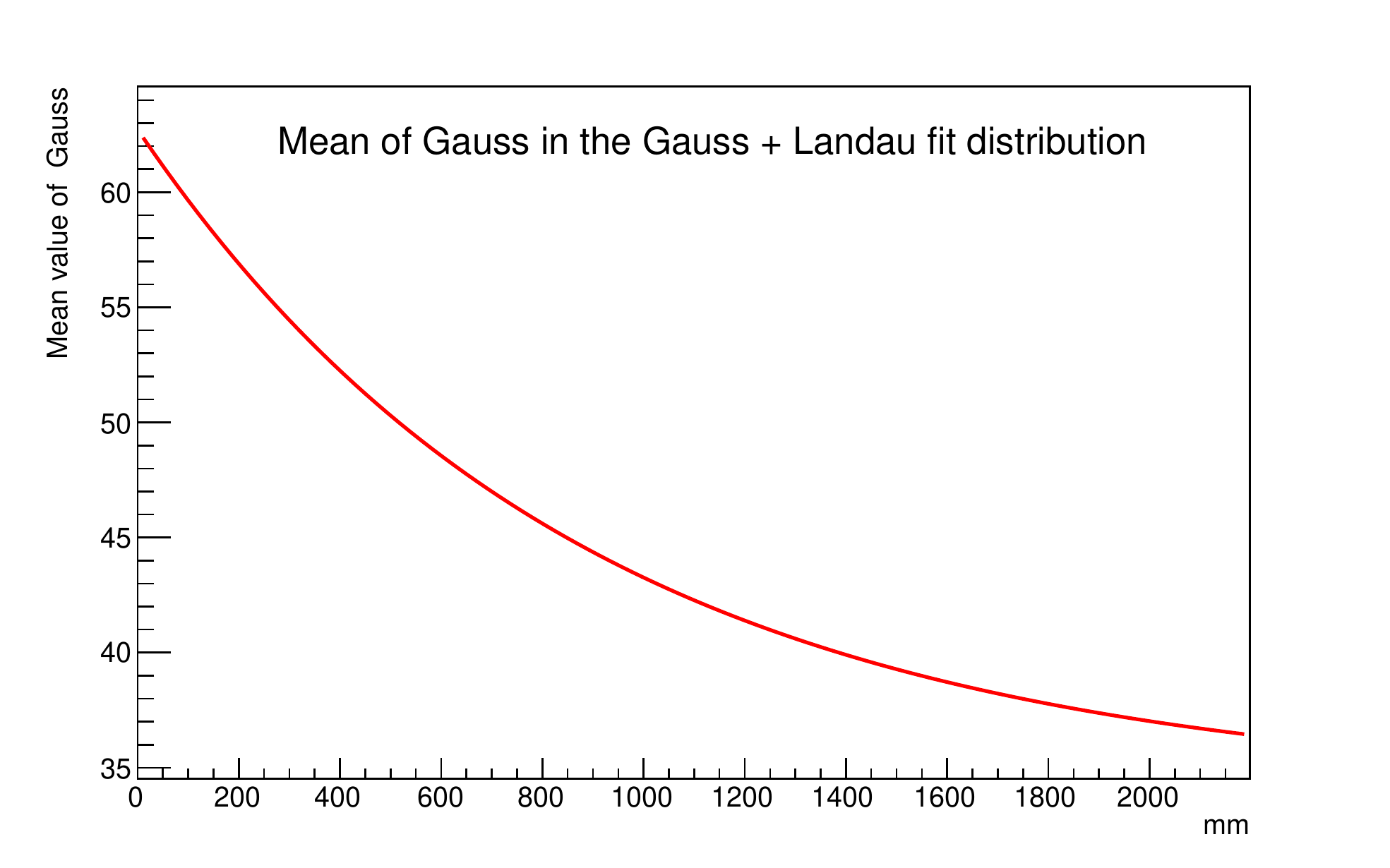}
\caption{ \textsf {Variation of a fit parameter of spectra along a 2218 mm strip.}}
\label{param}
\end{figure}

\begin{figure}
\centering
\includegraphics[width=\textwidth]{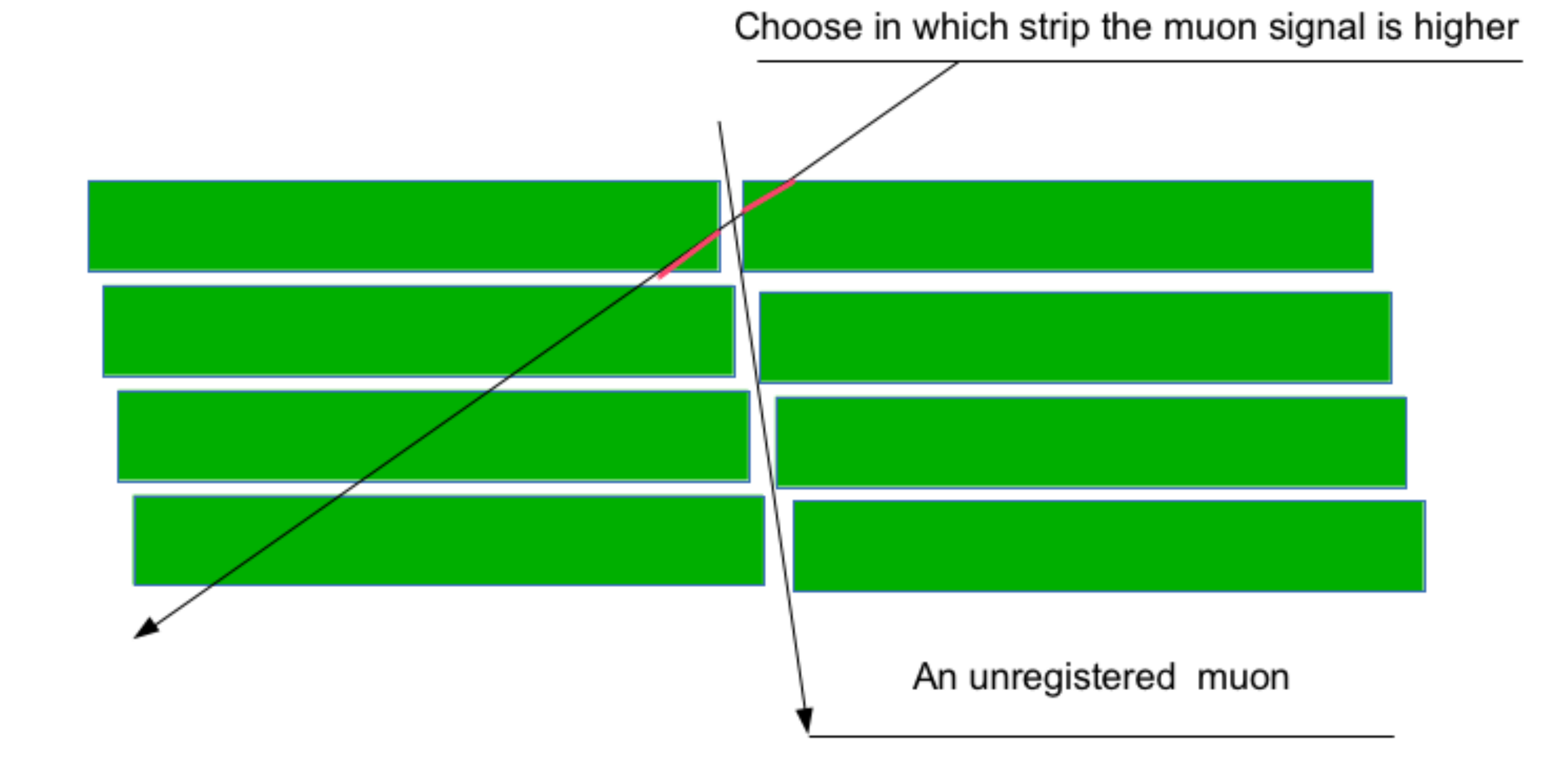}
\caption{ \textsf {An inter-strip gap.}}
\label{gap}
\end{figure}

Next steps were subsequently repeated for each layer until three of them gave signals below the threshold. 
\begin{itemize}
 \item interpolating an experimental spectrum of signals to the entrance point by means of fitting and interpolating fit parameters, see an example in fig.~\ref{param};
 \item calculating the range of muon in a hit strip, as well as in a neighboring one if it is penetrated by muon;
 \item the longest of the two ranges was used to scale the attenuated spectrum by means of scaling the fit parameters, cf. fig.~\ref{gap} ;
 \item comparing a random signal from the spectrum with the threshold value. 
\end{itemize}

\subsection{Simplifications}
\begin{figure}
\centering
\includegraphics[width=1.\textwidth]{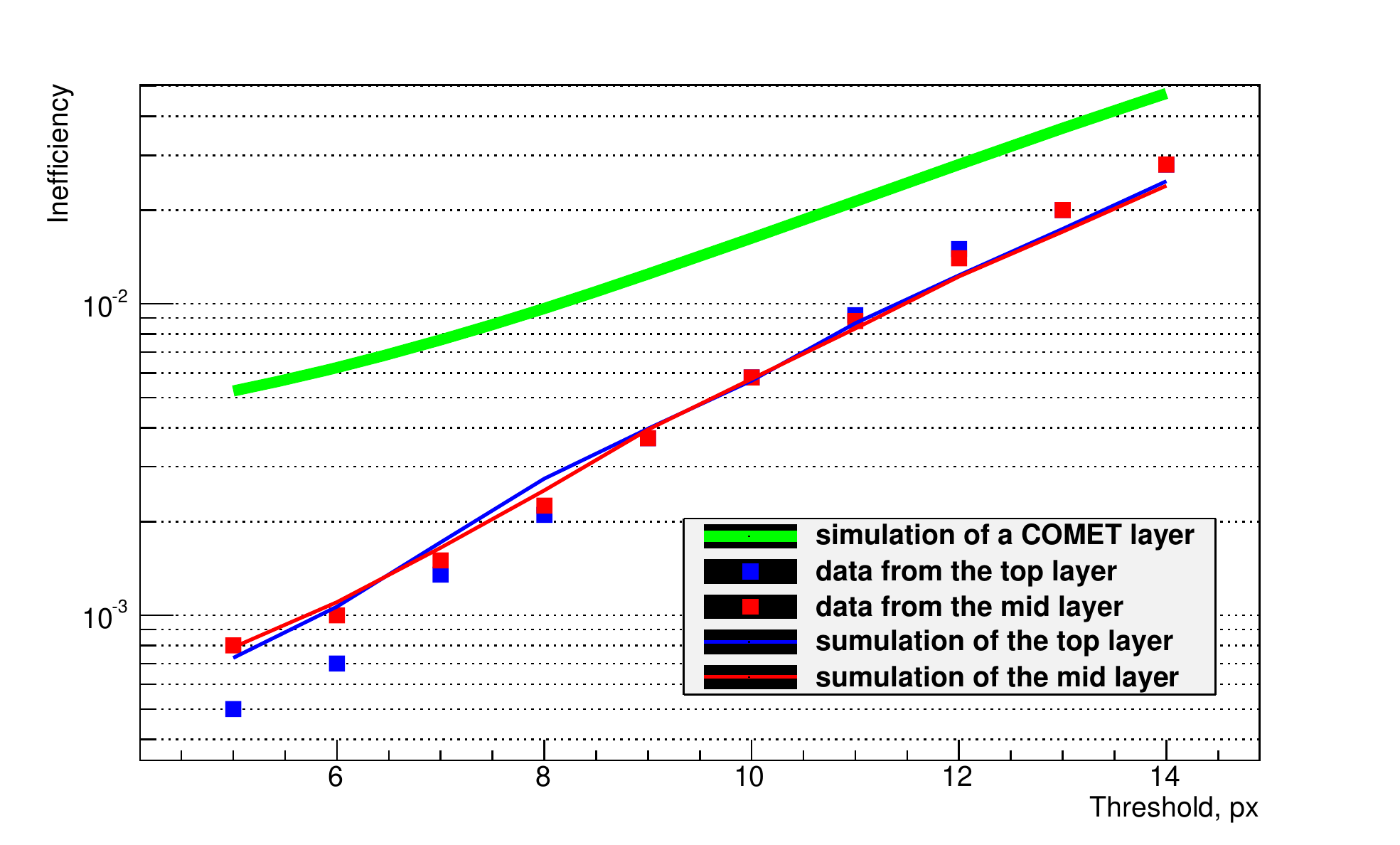}
\caption{ \textsf {Results of simulation  for  COMET single layer and for inefficiency measurement for both top and middle layers described in \cite{Tar_JNR}.}}
\label{ineff}
\end{figure}

For the sake of transparency of code and time saving, a number of simplifications of the real detector has been done in its model. In our opinion all those simplifications have a minor impact on results of simulation and  can therefore be treated more accurately  later on. Here is a list of them: 
\begin{itemize}
 \item only four modules of strips were  modeled for each layer;
 \item inactive zones of grooves for the WLS fiber were not modeled;
 \item crossing of more than two strips of a layer was neglected being both inessential and very rare for the given ratio of strip's thickness to width;
 \item a lateral distribution of the light in strips also was not modeled;   
 \item the experimental spectra were considered point-like while they were collected from a length 248~mm;
 \item the experimental spectra were fitted by a sum of Gauss and Landau distributions rather than by a convolution of the distributions;
 \item changing spectra with range was done by only scaling fit parameters while one expects widening  spectra for small ranges;
 \item to partially address the latter issue,  the Gauss distribution was replaced by the Poisson one in the case of a small range when a standard deviation of Gauss was smaller than a square root from its mean, which typically happened when the mean did not exceed  16~pixels;
 \item a coefficient  was empirically introduced to scaling of signals with range, which increases inefficiency and allows simulations of the inefficiency measurement \cite{Tar_JNR} fit results of that measurement, cf. fig.~\ref{ineff}. 
\end{itemize}

A straightforward, without the latter inter-calibration coefficient, simulation of that measurement  predicts smaller inefficiency.  The introduced coefficient  comprises  the effects of above simplifications as well as  the difference in responses between that measurement and obtaining the spectra at three locations used throughout the simulation.\footnote{ In this approach the value of simulated inefficiency must be conservative as the trigger in the inefficiency measurement had an admixture of ``empty'' events  of $10^{-3}$ order.}

\section{Results}
\subsection{Inefficiency}
Our simulation has shown that in a single COMET layer about one third of unregistered muons went through inter-module inactive zones. Others are not registered due to occasional small signals, predominantly at inter-strip \inzs\, in the rear part of strips where the attenuation of light is higher. Because of \inzs\, in the COMET layer, an average range of muons contributing to signal amounts 10~mm contrary to 10.5~mm in a continuous layer of scintillator. This takes place under the condition that only the largest  of two range parts contributes to the signal in the case when muons crosses an \inz. That condition a priori  increases the inefficiency of the \COl\, \wrt that of the measurement \cite{Tar_JNR} where both parts of range have been used -- cf. fig.~\ref{ineff} -- even though in that measurement the average range was only 9.8~mm because of its geometry.

\begin{figure}
\centering
\includegraphics[width=\textwidth]{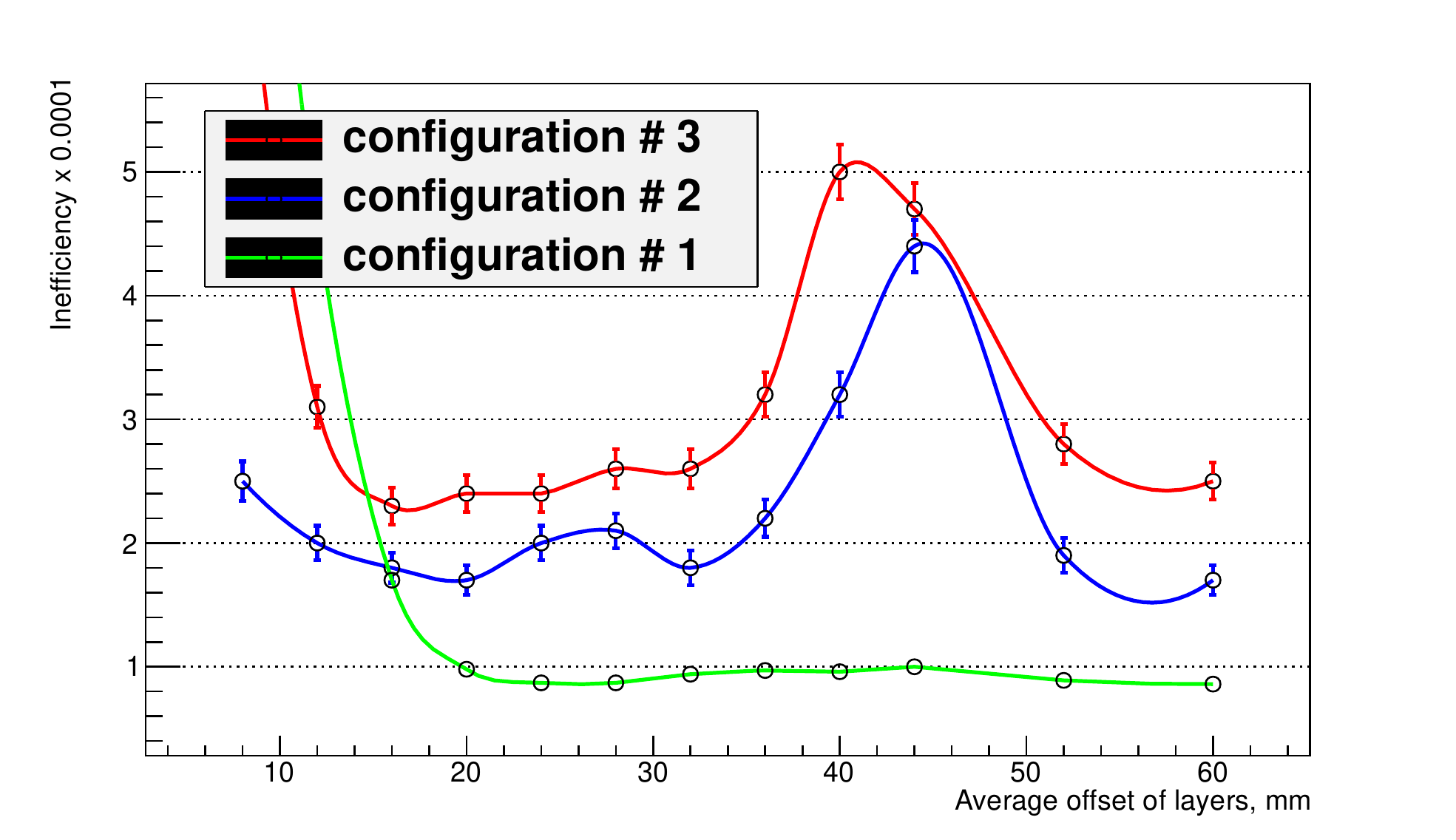}
\caption{\textsf{The muon registration inefficiency for a stack of four layers for the three configurations described in the beginning of the section~\ref{model}. Notice that the offset of the second layer in the first \con\, is a half of the \emph{average offset}.}}
\label{inef_four}
\end{figure}

As for the complete stack of 4 layers, in the second and the third \cons\, at the offset 40~mm positions of \ins\, \inzs\, of all layers coincide. At the offset  about 43~mm  the \ins\, \inzs\, of two layers appear below/above of \inm\, \inzs\, of other layers, right on the symmetry line of the zones. Such offsets abruptly increase the \ine, cf. fig.~\ref{inef_four}. The first \con\, is apparently preferable  because the inter-strip \inzs\, (i) do not coincide and (ii) can only be crossed by a muon trajectory at large polar angle where the muon flux is lower. The gap between layers of strips was omitted in the above simulations. One can expect to improve the performance of the that configuration by a sophisticated layout of \inzs. In the case of completely uncorrelated positions of \inzs\, in different layers the \ine\, should reach its low limit equal to four cubic inefficiencies of the single layer.

The muon-stopping target of the COMET detector is the most vulnerable element in the sense of cosmic muon interactions since the resulting electrons could not be geometrically distinguished from signal ones. To address this, for the first \con\, of the veto counter separate simulations of the registration \ine\,  for muons aimed to the target have been done. The \mst\, is located in the central region  of the detector beneath the center of the veto counter slab. Therefore it could be only hit by a small fraction of those muons that largely contribute to the registration \ine, i.e.coming through the \ins\, and \inm\, \inzs. This  makes the \ine\, lower. 

\begin{figure}
\centering
\includegraphics[width=\textwidth]{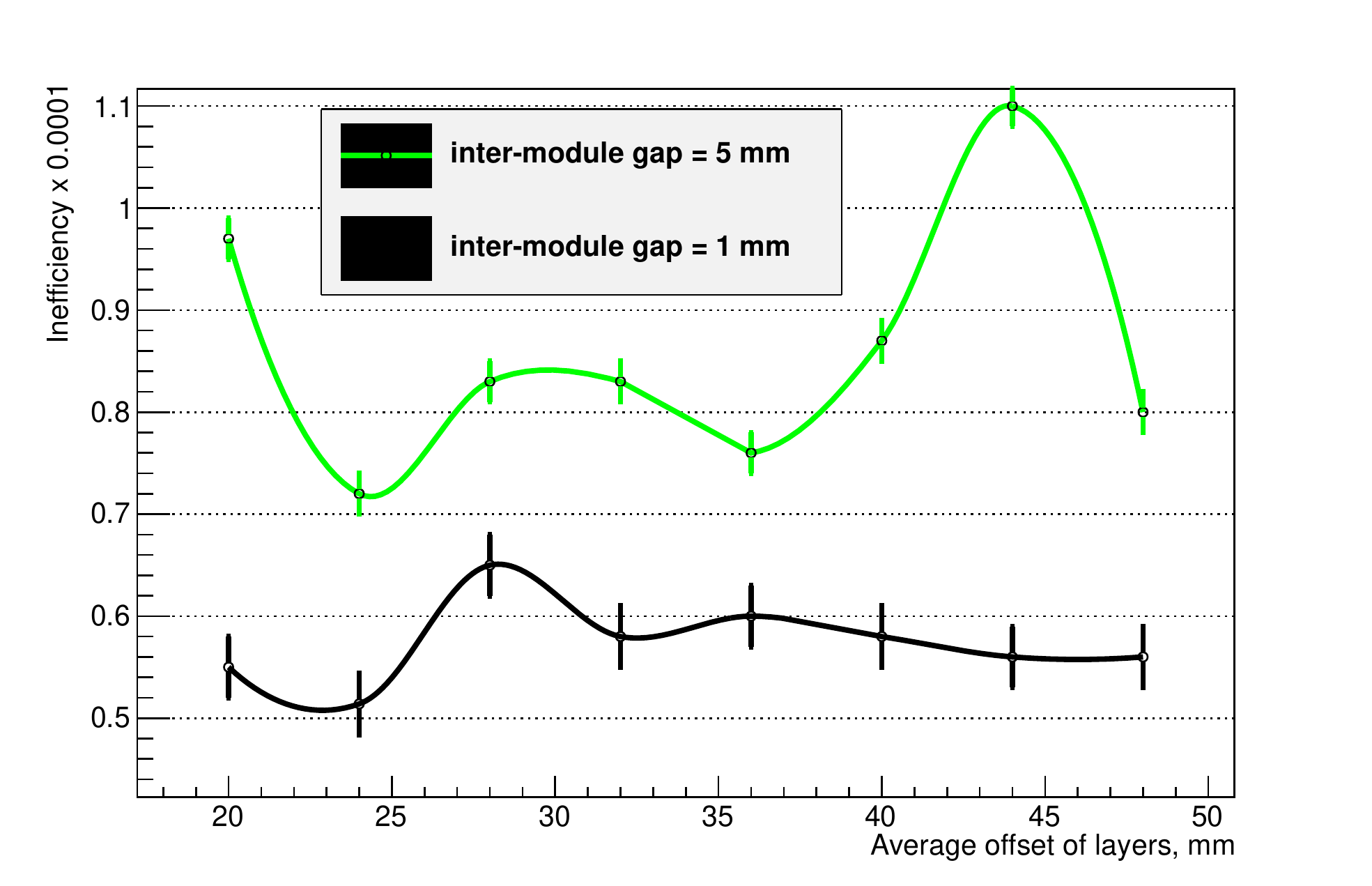}
\caption{\textsf{The muon registration inefficiency in the region of the \mst\, for two \inm\, gaps for the first configuration described in the beginning of the section~\ref{model}. The gap between the layers is taken to be 3 mm. Notice that the offset of the second layer is a half of the \emph{average offset}.}}
\label{target}
\end{figure}

To be more realistic, the constructing gap between layers equal to 3~mm was introduced in the simulations. Those simulations have been done for two different \inm\, gaps equal to 5~mm and 1~mm, cf. fig.~\ref{target}. The smaller \inm\, gap corresponds to the case of absence of the aluminum  module holder, which is a possible option since, unlike the Belle~II detector, the COMET veto counter  have no sloping modules to be held. In addition to the 11-pixel threshold, the simulations have been done for the readout thresholds equal to 10, 9, 8, and 7 pixels, which have shown the muon registration \ine\, decreasing by an order of magnitude when the threshold value was reduced by two pixels.    

\subsection{Noise signals}\label{noise}
The MPPC noise signals above the the threshold can fake muon passing, which results in the wasted time of data taking. The reason is the  random coincidence of MPPC noise signals that appear in  two strips of two layers within resolution time $\tau$, which could not be distinguished from  signals of a real muon in those layers. Veto time windows should be formed around such coincident noise signals.  Data from the veto windows should be excluded from the physics analysis, while in fact there was no any cosmic muons. As there is about two hundreds of strips in each layer of the horizontal slab of the veto counter, the frequency of the noise signal in the entire layer equals 200 times the frequency of the noise signal in a single strip $f$. 

During data taking that frequency grows with irradiation of photo-detectors by background neutrons that damage the p-n conjunction  of silicon photo-detectors \cite{Tar_Vienna}. The neutrons are abundantly  emitted by nuclei of aluminum of the muon-stopping target  after muon capture at the rate $\sim 10^{9}$ Hz. Signals of the background neutrons in the veto counter result also in the loss of data taking time. Fraction of the lost time is determined by the neutron registration efficiency and by neutron fluence at the veto counter. In order to minimize a frequency of neutron signals, the fluence of background neutrons will  be constrained   to $\sim 10^{5}$p/cm$^{2}$ by appropriate shielding. Therefore the irradiation dose received by MPPCs will not exceed $\sim 10^{12}$p/cm$^{2}$ throughout full time of COMET data taking.   

The frequency of the coincident noise signals in two layers can be  evaluated as  $200\cdot 2\tau f^2$  multiplied  by  the number of adjacent strips $k$  that could be crossed by a muon trajectory in other layers of the veto counter slab.  In order to cross underlying COMET detector elements, on which the cosmic muon could interact, the muon trajectory has to be descending, so for all practical purposes $k$ is not large: the steeper the trajectory descents, the smaller $k$ is.  

For instance to cross the muon-stopping target, the $\theta$-angle of muon trajectory could not exceed about 50\textdegree \, for the counter dimensions quoted in \cite{Drutsk}. Taking into account the value of the angle and the  ratio of strip's thickness to width, one can see that after crossing the first layer the trajectories aimed to the target can only cross one of three adjacent strips in the second layer, one of five strips in the third one, and one of seven strips  in the fourth one, i.e. $k=15$. For the same reasons a muon signal in some strip of the second layer can only coincide  with a signal from one of $3+5$ adjacent strips of two subsequent layers, similarly a signal in the third layer can coincide with that in three adjacent strips of the fourth layer. Summing up the probabilities for all layers one gets the increasing factor 26, thus the frequency  of  the fake muon signal  is about $10400 \tau n^2$.

\begin{figure}
\centering
\includegraphics[width=\textwidth]{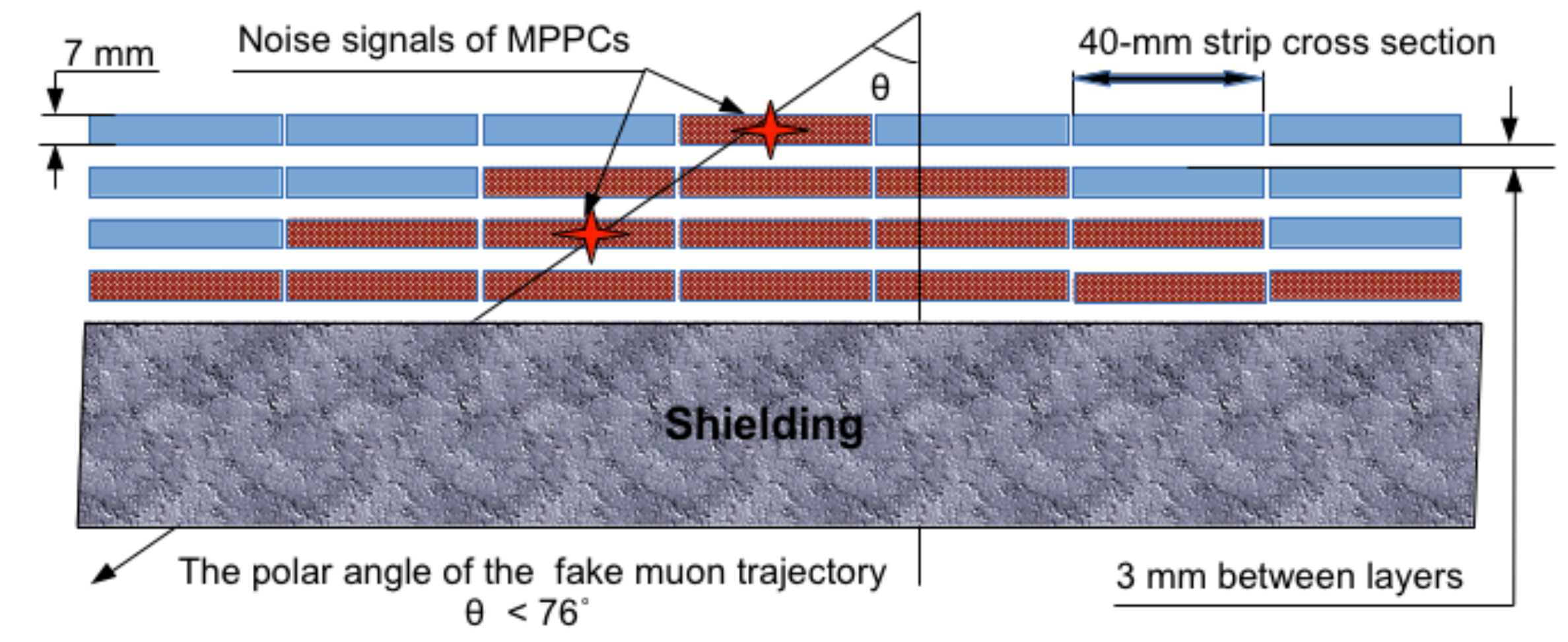}
\caption{\textsf{The minimal scheme of strips that should be examined for a coincident noise signal.}}
\label{count}
\end{figure}

\begin{figure}
\centering
\includegraphics[width= \textwidth]{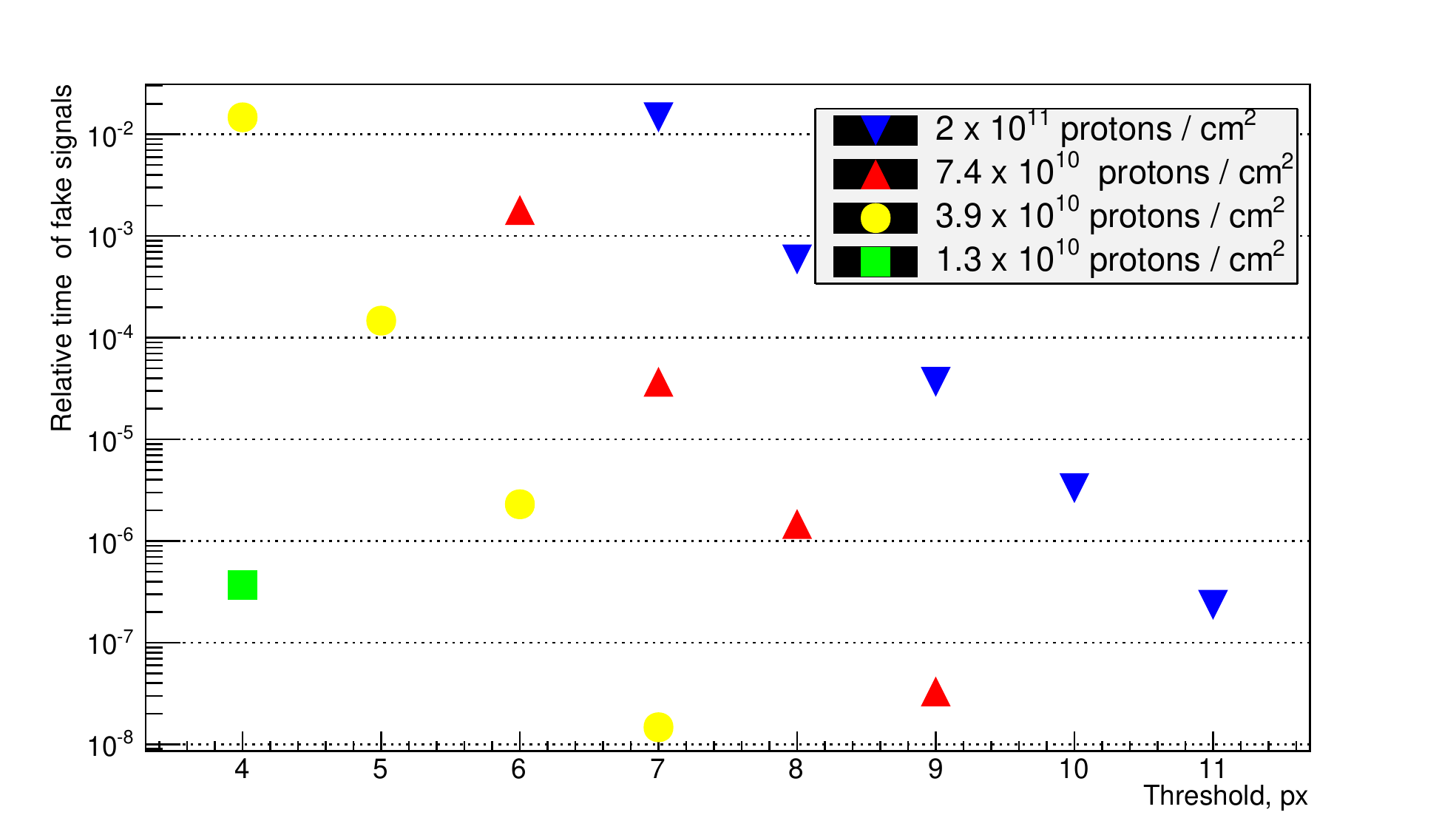}
\caption{\textsf{The fraction of time lost due  to fake muon signals caused by coincident noise signals of MPPC vs. the readout threshold for four irradiation doses of MPPCs: $1.3 \cdot 10^{10}$ p/cm$^2$, $3.9 \cdot 10^{10}$ p/cm$^2$, $7.4 \cdot 10^{10}$ p/cm$^2$, and  $2 \cdot 10^{11}$ p/cm$^2$ (based on \cite{Tar_Vienna}). The length of the veto window is taken to be 100~ns, the signal resolution time of the veto counter being 20~ns.}}
\label{fake}
\end{figure}

In the general case, when one looks for the coincident signal in $1+2s$ adjacent strips of the second layer, $k=3+12s$ and the frequency-increasing factor equals to $6+20s$.  For quoted strip dimensions and the  3 mm gap between layers, that  minimal value of $k=15$ allows covering of the   $\theta$-angle up  to  76\textdegree, cf. fig.~\ref{count}.
For $s=2$ one covers the region up to 83\textdegree of $\theta$, which is sufficient to survey the whole detector if the latter is separated from the veto counter by 500 mm neutron shielding. The 125~mm distance between the counter and the detector assumes examination of 99 strips covering 88\textdegree. Figure~\ref{fake}  shows the fraction of time lost due to fake veto signals versus the readout threshold,  for different irradiation doses received by MPPC, and  $s=2$.

\section{Conclusion}
A simplified computer model of the scintillator strip COMET veto counter demonstrates  viability of  the counter \ine\, to be 0.0001. It has been shown, that operating at the threshold above seven pixels can keep the fraction of the lost time at the level of few percents. A disadvantage of working  at high thresholds is lower muon registration efficiency.  However, simulations of the efficiency value has demonstrated that it still meets COMET requirements  even at the 11-pixel threshold. On the other hand both the noise frequency and the neutron registration efficiency are lower when operating at high thresholds, which results in a shorter time loss.

\begin{figure}
\centering
\includegraphics[width=1.\textwidth]{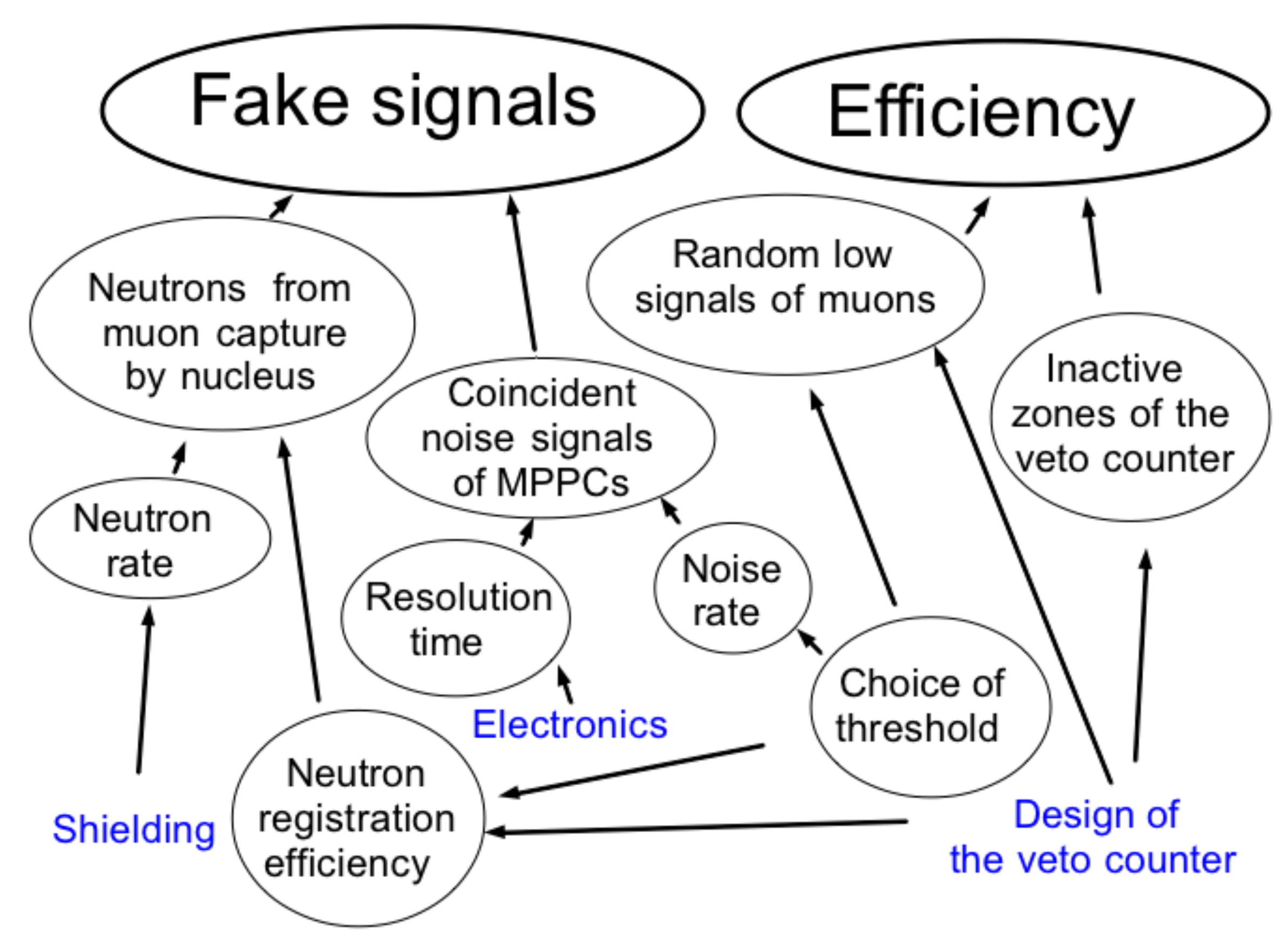}
\caption{ \textsf {Dependencies for optimization  of the veto counter.}}
\label{logic}
\end{figure}

Further work should be done in the two basic directions: (i) building an elaborate model of the veto counter and (ii) optimization of the counter design according to the scheme in fig.~\ref{logic}. We are grateful to B.~Bobchenko, A.~Drutskoy and V.~Rusinov for discussions and help.

\end{document}